\documentclass[letter]{jpconf}
\usepackage{graphicx}

\newcommand{\gev}{\,\mathrm{GeV}}
\newcommand{\mev}{\,\mathrm{MeV}}
\newcommand{\kev}{\,\mathrm{keV}}

\begin{document}
\title{Hadronic Physics}

\author{Hanna Mahlke}

\address{Laboratory for Elementary-Particle Physics, Cornell University,
Ithaca, NY 14853, USA}

\ead{mahlke@mail.lepp.cornell.edu}

\begin{abstract}
A selection of studies highlighting different manifestations
of the strong interaction are presented. Many new results
have become available this summer in the regimes of
discovery, systematic survey, and precision measurements
of bound quark states.
\end{abstract}

\section{Introduction}
The investigation of bound states of quarks provides insight 
into the strong force, by which they are held together. Doing so
with different kinds of hadrons allows one to study different
manifestations of QCD. Especially at low scales, however, 
modeling such phenomena is challenging. 

Quark models predict systems of states for different quark
configurations that can be confronted with experimental data.
While several possible configurations would in principle
be useful to study, including bound gluon-only states,
the objects that lend themselves to this exercise more
easily are mesons and baryons. 
Beyond these two-quark or three-quark systems predicted within
this framework, other bound configurations can exist that
have quantum numbers not predicted in the nonrelativistic
quark model (``exotics'') or have four or five quark constituents.

This summary focuses on results that were new this summer.
A selection of topics will be presented, illustrating 
ways to study the effects of strong interaction from low to
high energies. This covers baryons and mesons.
For a summary on pentaquarks, I refer the
interested reader to Ref.~\cite{pentaquarks}.
The analyses divide into three categories:
(1) new phenomena, (2) systematic surveys and complementary
approaches to verify new phenomena, (3) precision measurements.

\section{Two heavy quarks}
Heavy quarkonium, bound states such as $c \bar c$ or $b \bar b$,
are an ideal laboratory for studing the strong potential as the 
masses and widths of states are directly related to the strong 
force holding them together, mediated by gluons. 
From a researcher's standpoint, this is the same scenario as 
positronium, where the electron and positron are
held together by the electric force, mediated by photons.
However, the energy scale is vastly different for hadrons.
Bottomonium, with heavier constituents of about $5\gev$
than charmonium, can be treated as non-relativistic. 
On the other hand, charmonium, 
with the lowest-lying state at just below $3\gev$,
affords the opportunity to study the importance
of relativistic corrections. 
I will list examples for spectroscopy, decay, 
and searches for quarkonium states. While substantial
bottomonium samples on the $\Upsilon(1,2,3S)$ exist
(direct production or radiative return from higher
energies), most progress this summer was made for charmonium.

\subsection{Below open flavor threshold}
All charmonium states below open-flavor threshold 
have been observed~\cite{pdg06}. 
Large samples exist for the $J/\psi$ and $\psi(2S)$,
which are well-studied. 
The masses and widths have been determined at high accuracy.
The focus is now on comparing the two states, identifying rare decays, 
and investigating the resonant substructure in multibody states. 
As for open charm,
this provides information on the intermediate states produced
and gives insight into the decay dynamics.
A scan of the $\psi(2S)$ by E385~\cite{e385:psi2Swidth}
led to the current most precise results of 
$\Gamma(\psi(2S))=(290 \pm 25 \pm 4)\kev$ and 
$\Gamma(\psi(2S) \to e^+e^-) \times {\cal B} (\psi(2S) \to p \bar p)
=(0.579 \pm 0.038 \pm 0.036) \kev$.
The process used was $p \bar p \to \psi(2S)$ with
$\psi(2S) \to e^+e^-$ or $\psi(2S) \to X J/\psi \to X e^+e^-$.
The analysis makes use of the small beam
energy spread, which is comparable to the structure investigated
as opposed to the $\sim~$MeV range to which $e^+e^-$ machines
are limited.

The $\chi_{cJ}$ states can be studied using the reaction
$\psi(2S) \to \gamma \chi_{cJ}$, where they are produced
at a branching ratio of a little under $10\%$ each. 
Once the transition photon is identified, 
the $\chi_{cJ}$ are easy to handle experimentally.
Given the
$\psi(2S)$ sample sizes, this implies that the $\chi_{cJ}$
data are not far behind the $\psi(2S)$ in statistical power, 
and similar studies as for the $\psi(2S)$ are being conducted.
The transition rates are affected by relativistic corrections,
and thus measuring them accurately is important to guide theory.
The $\eta_c(1S,2S)$ and the $h_c$ are less well known, and studies
to learn more about their properties and decays are under way.

CLEO-c presented a study of decays 
$\chi_{cJ} \to h^+h^-h^0\pi^0$~\cite{cleo:chicJ_4h}.
Branching fractions for the final states
$\pi^+\pi^-\pi^0\pi^0$ (and resonant sub-mode $\rho^\pm \pi^\mp \pi^0$),
$K^+K^-\pi^0\pi^0$ (and resonant sub-mode $K^* K\pi$),
$p \bar p \pi^0\pi^0$,
$K^+K^-\eta\pi^0$, and
$K^\pm \pi^\mp K^0 \pi^0$,
were determined. Most of these are first measurements. 
Isospin relations in the submodes $\rho^\pm \pi^\mp \pi^0$ and $K^* K\pi$
are found to be consistent with expectations.

BES investigated the decay $J/\psi \to K^+K^-\pi^0$~\cite{bes:X1576}.
The distribution
of $M^2(K^+\pi^0)$ {\sl vs.} $M^2(K^-\pi^0)$ shows the expected bands
of the $K^*(892)$ and the $K^*(1410)$. 
The necessary fit components
to get a reasonable description of the data are $K^*(892)$,
$K^*(1410)$, $\rho(1700)$, a non-resonant contribution, and
a Breit-Wigner with a mass-dependent width
function for a broad enhancement in $M(K^+K^-)$ around $1.6\gev$, 
which does not match any known particles. It is not
possible to use $\rho$ excitations to describe the observed
distribution. 
$C$-parity conservation implies that this state should have
odd $J$ and $PC=--$; the fit prefers $1^{--}$. 
The pole position, which is chosen instead of 
Breit-Wigner-like parameters to quote a result,
as determined by the fit is
  $(1576 ^{+49 +98}_{-55-91})\mev 
-i (409^{+11+32}_{-12-67})\mev$; a product branching
fraction ${\cal B}(J/\psi \to X \pi^0) \times
{\cal B}(X \to K^+K^-) 
= (8.5 \pm 0.6^{+2.7}_{-3.6} ) \times 10^{-4}$
is found.
Future studies will focus on a search for the same phenomenon
in isospin-related final states to clarify the nature
of this state.

Studies of charmonium production in $B$~decay
aim at gaining insight into the production
mechanism and variations among different $c\bar c$ states.
The $\eta_c$, $J/\psi$ and $\chi_{c1}$ (and excitations) along with
a $K$ or $K^*$ can be created through $b \to c \bar c s $, 
but production of $h_c$, $\chi_{c0}$, and $\chi_{c2}$ must 
involve different mechanisms. A prediction~\cite{bodwin:CcbarInBDecay} 
states that they can be produced as copiously as
$\chi_{c1}$ (the branching fractions ${\cal B} (B \to \chi_{c1} K^{(*)}$
are of order $10^{-4}$). 
This holds for the $\chi_{c0}$, at ${\cal B}(B^+ \to \chi_{c0} K^+)
= (1.4 ^{+0.23}_{-0.19}) \times 10^{-4}$,
but the current upper limits for $h_c$ (from $B^+ \to h_c K^+$) 
and $\chi_{c2}$ are an order of magnitude lower.
BaBar presented preliminary results for $B^+$ and $B^0$ decay
to final states with $\eta_c \to K_S K^+ \pi^-, \ K^+K^-\pi^0$
and $h_c \to \gamma \eta_c$~\cite{babar:BToh_cAndEta_c}:
${\cal B} (B^0 \to \eta_c K^{*0}) = ( 6.1 \pm 1.4 )\times 10^{-4}$ 
         (uncertainty improved by 50\%),
${\cal B} (B^+ \to h_c K^+) \times {\cal B}(h_c \to \gamma \eta_c) 
         < 5.2 \times 10^{-5}$ at 90\%~CL (in agreement with Belle),
${\cal B} (B^0 \to h_c K^{*0}) \times {\cal B}(h_c \to \gamma \eta_c)
         < 2.4 \times 10^{-4}$ at 90\%CL (first limit).
The branching fraction ${\cal B}(h_c \to \gamma \eta_c)$ is not
known. The current level of sensitivity does not yet allow 
a firm conclusion on the level of
suppression of $h_c$ production.

\subsection{Above open flavor threshold}
Little is known about charmonium states above $D \bar D$ threshold.
Candidates for the states
$3^3S_1$, $2^3D_1$, and $4^3S_1$ ($J^{PC}=1^{--}$) are identified as
peaks in the spectrum of the inclusive hadronic cross-section;
their positions and widths match theoretical predictions for those states. 
Observation of other states not accessible in $e^+e^-$ collisions
is possible in $B$~decay,
$p \bar p$ production, or through a transition from a higher-mass
state.

A fit to the $R$ spectrum provides the masses and widths of the
$\psi(4040)$, $\psi(4160)$, $\psi(4415)$, but the extraction of
these quantities is not without ambiguity. 
BES, for the first time,
attempted to take interference between these broad resonances
into account~\cite{bes:2-5GevScan}. 
The parameters determined show substantial variation
with respect to a fit without interference.

An interesting question is what the inclusive cross-section
is composed of; a question that a fit to the cross-sections
measured for exclusive decay samples in data taken in the 
charm region (CLEO) or with initial state radiation from higher
energies (BaBar, Belle) can answer. Belle presented a new 
study~\cite{belle:DD2star}
of $D \bar D$ and $D \bar D \pi$ (not through a $D^*$) that 
augmented an earlier publication on 
$D \bar D^*$, $D^*D^*$~\cite{belle:ddstar_dstardstar}. 
Summed up, 
the features of the inclusive spectrum from BES are reproduced,
aside from a shift due to the smooth contribution from $uds$ continuum.
The components not present in such a comparison are expected to
be of comparatively
low cross-section: charmonium production, $D_s$ production,
charm baryons, other $D \bar D n\pi$ (non-resonant), 
$D \bar D^* \pi$ (CLEO \cite{cleo:dstardpi}).

The distribution of $D \pi$ in the $D \bar D \pi$ sample shows
a preference for $M(D^- \pi^+)$ near the $D_2^*(2460)$. Selecting
events of this type, namely $D D_2^*(2460)$, and plotting their
invariant mass, a peak of $~14\sigma$ statistical significance
at the $\psi(4415)$ is found at 
$m=(4.411 \pm 0.007 (\mathrm{stat}))\mev$,
$\Gamma=(77 \pm 20 (\mathrm{stat}))\mev$. 
The mass and width extracted match
those from the inclusive BES analysis~\cite{bes:2-5GevScan}, 
$m=(4.4152 \pm 0.0075 (\mathrm{stat}))\mev$ and
$\Gamma=(73.3 \pm 21.2 (\mathrm{stat}))\mev$.
The $D \bar D \pi$ events where $M(D^- \pi^+)$ is outside the 
$D_2^*(2460)$ region are consistent with background from sidebands
below $4.6\gev$ and show a slow rise thereafter,
consistent with CLEO-c's findings~\cite{cleo:dstardpi}, 
where $D \bar D \pi$
is not observed at energies below $4.26\gev$.

No other exclusive decay branching fractions have been measured
for charmonium states above the $\psi(3770)$ 
(only upper limits exist)~\cite{pdg06}.

The properties such as production / decay patterns or 
masses / widths of some other states have been found to
resemble those of expected charmonium states.
In most cases, it is difficult
to come up with an unambiguous assignment for them
based on the experimental evidence available to date.
Examples follow.

Belle analyzed their ISR data for the decay 
$e^+ e^- \to \gamma (\pi^+ \pi^- J/\psi)$~\cite{belle:y4260}, 
in order to improve knowledge of the $Y(4260)$.
This state has been seen before by BaBar, Belle, and CLEO,
but thus far only been fit to a single Breit-Wigner 
resonance. Belle attempted a fit using two Breit-Wigner
shapes that interfere so as to obtain a better description
in particular of the low-side tail of the $Y(4260)$. 
Due to a mathematical ambiguity, two solutions are found that 
yield an identical description of the data,
have the same values for masses and width, but
result in substantially different product of couplings
of the $Y(4260)$ and the $X(4008)$ (the resonance
associated with the low-side Breit-Wigner) to the 
initial and final state, 
${\cal B}(X \to J/\psi \pi^+\pi^-) \times \Gamma_{ee}$. 
The properties of the $Y(4260)$ contribution are 
consistent with those published earlier by BaBar and CLEO.

The distribution of $m(\pi^+\pi^-)$ favors higher values
for events taken from $m(\pi^+\pi^- J/\psi)$ near the
$Y(4260)$, matching the observation by other experiments,
but is consistent with phase space for events
above or below the $Y(4260)$ peak region.

Belle also searched for 
$e^+ e^- \to \gamma (\pi^+ \pi^- \psi(2S))$~\cite{belle:psi2Spipi}. 
Belle confirmed the BaBar observation of a peak at 
$m(\pi^+ \pi^- \psi(2S)) = 4.35\mev$, with similar
parameters, but a second peak at $4.66\gev$ is found:
$m=(4664\pm 11\pm 5)\mev$,  $\Gamma = (48\pm 15\pm 3)\mev$.
The distribution of $m(\pi^+\pi^-)$ is inconsistent with
phase-space decay for the lower-mass peak, but shifted towards
higher values. In the case of the second peak, the distribution
strongly favors high values; the $m(\pi^+\pi^-)$ distribution
is suggestive of the $f_0(980)$.

Another as yet unexplained state is the $X(3940)$,
observed by Belle in $e^+e^- \to J/\psi D^{(*)} D^{(*)}$
as a peak in the $m(D^*D)$ distribution of events recoiling
against the $J/\psi$. More Belle data confirm~\cite{belle:x4160}
the existence of this state, with consistent parameters but improved
significance. A question that needs to be addressed in
order to facilitate a quantum number assignment and
placement in the charmonium system is to which degree this 
state decays to $D \bar D$, and to also check for structure in
$D^* \bar D^*$ ($m(X(3940)$ is well below $2m(D^*)$ though). 
The $m(D \bar D)$
spectrum shows no peak at $3.94\gev$ above background, 
but a broad threshold enhancement. Before a limit for
${\cal B}(X(3940) \to D \bar D)$ can be set, this will have to be
understood. In $m(D^*D^*)$, a broad peak of more than five sigma
statistical significance is observed, which is identified
as a new particle $X(4160)$ and fit with a Breit-Wigner function.
While the parameters for the $X(4160)$ differ from those
of the $X(3940)$, there is some overlap due to the large
widths of both.

BaBar confirmed Belle's observation~\cite{belle:y3940} 
of the $Y(3940)$ in $B \to K (\omega J/\psi)$, with substantially
improved statistical and systematic uncertainties~\cite{babar:y3940}. 
The BaBar mass and width are substantially, but not significantly
($-1.7\sigma$ and $-1.5\sigma$, respectively), lower than 
Belle's.

\section{One heavy quark}
The theoretical treatment of particles with one heavy and one
light quark differs in that the two degrees of freedom are
decoupled and the heavy quark can be treated as stationary.
Similar guidelines as in quarkonium apply,
and it is important to aim for a complete picture in which
the existence and properties of the expected states are 
searched for.

Belle investigated angular distributions 
of the decay $D_{s1}(2536)^+ \to D^{*+}K_S$~\cite{belle:ds1_2536}, 
a state which has been observed before: $J^P=1^+$,   
total angular momentum of the light degrees of freedom
$j=3/2$, known decays $D^{*+}K_S$, $D^{*0}K^+$,
$D_s^+\pi^+\pi^-$. 
A new mode was observed,
$D_{s1} \to  D^+ \pi^- K^+$ (but not through $D^{*0}$), 
and the ratio 
${\cal B}(D_{s1}^+ \to D^+ \pi^- K^+) 
/
 {\cal B}(D_{s1}^+ \to D^{*+} K^0)$
is found to be $(3.17 \pm 0.17 \pm 0.36)\%$ (preliminary).
The angular study of $D_{s1} \to  D^* K_S$
provides a handle on the mixing between two 
$J^P = 1^+$ states $D_{s1}$ and $D_{sJ}(2460)$. 
Within the Heavy Quark Symmetry the state with 
$j=3/2$ 
(which is naively expected to be $D_{s1}$) 
decays to $D^* K_S$ in a pure $D$-wave, while the one with
$j=1/2$ does so in an $S$-wave. Mixing between the two
can result in an $S$-wave component in $D_{s1}(2536)$ decay.
This contribution
to the total width is determined from a partial wave analysis 
and found to be substantial: $\Gamma_{S\mathrm{-wave}}
/ \Gamma_{\mathrm{total}} = 0.72 \pm 0.05$ (preliminary).

Much more progress has been made in meson and baryon
spectroscopy~\cite{baryon:spectroscopy,babar:baryons}.
BaBar conducted a systematic search for charm baryons decaying
to final states $\Lambda_c^+$ plus 
$K_S$, $K^-$, $K^- \pi^+$, $K_S \pi^-$, $K_S \pi^+\pi^0$, 
$K^-\pi^+\pi^-$~\cite{babar:baryons}.
Their preliminary results confirm the 
states $\Xi_c(2980)^+$,
$\Xi_c(3077)^+$, and $\Xi_c(3077)^0$, improve the mass
and width measurements of $\Xi_c(2980)^+$, and discover
two new states:
$\Xi_c(3055)^+$ and $\Xi_c(3123)^+$. They are only
observed by their decays to $\Sigma_c(2455)^{++}K$
and $\Sigma_c(2520)^{++}K$, implying that contrary to
the other cases the $c$ and the $s$ quark in the parent
particle separate.

\section{Zero heavy quarks}

Theoretical treatment of mesons with only light quarks demands
non-perturbative methods. Guidance in these soft processes comes
from scattering experiments as well as studies of decays.

KLOE investigated the decay 
$\phi \to \pi^0 \pi^0 \gamma$~\cite{kloe:pi0pi0gamma} to
help shed light on the nature of the $f_0(980)$.
The analysis complements that of other final states such
as $\pi^+ \pi^- \gamma$ or $\eta \pi^0 \gamma$ that are
aimed at the $f_0(980)$ and $a_0(980)$,
broad scalar resonances that appear as intermediate states
also 
in many heavy-quark decays. Producing them in $\phi$ decay
makes it possible to study them close to their production
threshold. The Dalitz plot $m(\pi^0\gamma)^2$ {\sl vs.}~$m(\pi^0\pi^0)$
is fit with two different models that test for the existence of
an intermediate kaon loop ($\phi \to \gamma K^+ K^- \to
f_0(980) \to \pi^0\pi^0$) or a pointlike coupling
($\phi \to f_0(980) \to \pi^0\pi^0$). Both models fit the
data resonably well, couplings are measured,
and the product branching ratio is determined.

The $\eta$-$\eta'$ system
is often parametrized by a mixture of two components:
$|u \bar u + d \bar d \rangle / \sqrt 2 $
and $| s \bar s \rangle$. 
The mixing angle can be determined for instance from
the ratio ${\cal B}(\phi \to \eta'\gamma) 
/ {\cal B}(\phi \to \eta \gamma)$. 
Under the assumption that no gluonium contributes,
KLOE determines the mixing angle in the quark-flavor basis to be 
$\phi_P = (41.4 \pm 0.3 \pm 0.7 \pm 0.6)^\circ$~\cite{kloe:gluonium},
the most precise result to date. 
States of pure glue content are predicted to exist
at higher energies, yet low enough that they may mix
with the $\eta'$. Allowing for such a component in the $\eta'$
introduces another mixing angle to quantify the
gluonium contribution. Within this parametrization
and combined with other such ratios from external
input, an improved agreement between SU(3) predictions
and the observed branching fraction results
is achieved if a gluonium component is allowed.
The squared amplitude coefficient for this component 
is determined to be $(14 \pm 4)\%$.
The result for $\theta_P$ is not very sensitive to this
change in parametrization.
The range of values for $\theta_P$ determined here
is consistent with determinations in the 
octet-singlet basis in other experiments.

Searches for decay to undetectable final states
are not only relevant in the perpetual quest for new
physics, but also so as to ensure that the total decay 
width may be approximated by sum of all observed decays. 
BES searched for the decay $\eta(') \to
\mbox{undetectable final states}$, where the $\eta(')$ is 
produced in the reaction 
$J/\psi \to \phi \eta(')$~\cite{bes:etainvis}. 
The $\phi$ as a narrow resonance
is readily identified via its decay into a charged kaon pair, and kinematics
constrain the recoiling $\eta(')$ to a narrow region in the missing momentum.
No signal is seen; an upper limit is placed on the decay of $\eta(')$
to invisible final states relative to decay into two photons, which translate
into absolute branching fractions of $\eta,\ \eta' \to \mbox{invisible}$
of $7 \times 10^{-4}$ and $2 \times 10^{-3}$, respectively. 
For comparison, similar
studies for the $\Upsilon(1S)$ led to a limit of $0.3\%$~\cite{pdg06}.

CLEO used the transition $\psi(2S) \to \eta J/\psi$ with
$J/\psi \to \ell^+\ell^-$ 
to study the $\eta$ meson.
Branching fractions and ratios thereof~\cite{cleo:etabr} were determined for
$\eta \to \gamma\gamma$, $\pi^+\pi^-\pi^0$, $3\pi^0$, $\pi^+\pi^-\gamma$,
and $e^+e^-\gamma$, a first for such a suite of modes within the same
experiment. These branching fractions are of order $1\%$ or larger;
the ones not covered are at least an order of magnitude lower, and
their sum is estimated to amount to no more than $0.2\%$. 
Deviations from previous determinations were observed for
$\pi^+\pi^-\gamma$ and $e^+e^-\gamma$ at the level of three standard
deviations. The kinematic conditions allowed CLEO to measure the
$\eta$ mass~\cite{cleo:etamass}
through a fit of the invariant mass of the decay products (except
$e^+e^-\gamma$, which has too few events). The precision achieved is
comparable with that of dedicated experiments. CLEO, agreeing with 
NA48 and KLOE, indicates a recent GEM result as an outlier.

KLOE's high-statistics
study~\cite{kloe:eta_3pi} of $\eta \to \pi^+\pi^-\pi^0$
aims at testing the degree to which lowest-order current algebra
is able to describe the decay dynamics. If this is accurate,
then the decay amplitude can be used to extract a measurement of
the $u$-$d$ quark mass difference in a simple way. 
A fit to the Dalitz plot is performed up to third
order in kinematic quantities. While coefficients that would indicate
charge violation are found to be zero (current best limits), 
in line with expectation,
comparing a relationship between other components reveals that
lowest-order current algebra is not sufficient.

The decay rate for $K^+ \to \pi^+ \pi^0 \pi^0$ can be used
to understand final state interactions, undertaken by
NA48~\cite{na48:k_3pi_and_ke4}. In a simplified picture where
the production of the three pions is instantaneous,
the $m(\pi^0\pi^0)^2$ distribution shows a rapid rise
at the kinematic limit and then quickly changes slope
to an almost linear behavior. Experiment shows the
rise to be slower than expected 
and a distinct change in behavior (``cusp'') 
at $\pi^+\pi^-$ production threshold. 
Below this point, there is a depletion of events relative
to the expectation.  This is
due to the fact that the amplitude for the 
direct decay $K^+ \to \pi^+ \pi^0 \pi^0$
interferes with rescattering amplitudes, for example the
one-loop process 
$K^+ \to \pi^+ ( \pi^+ \pi^-) \to \pi^+ \pi^0 \pi^0$.
The area above the cusp allows to observe sub-leading
effects.  NA48's high-precision data (updated, preliminary) 
allow to explore the sensitivity of the Dalitz plot
to the scattering lengths $a_0$ and $a_2$, for which 
theoretical predictions from chiral perturbation theory exist.
The measured values are consistent with those obtained in 
$K \to \pi^+\pi^- e \nu$ and pionium lifetime measurements.

A clean way to explore final state interactions close
to threshold is $K \to \pi^+\pi^- e \nu$, with new work
by NA48~\cite{na48:k_3pi_and_ke4}.
The fit to describe the amplitude
uses a model-independent approach to measure 
form factor coefficients, achieving a new level of sensitivity,
and allows to extract $a_0$ and $a_2$ in an
independent manner (albeit with further theoretical
input). The new preliminary results are consistent with
an earlier publication on a partial sample.
For both the $K \to \pi^+\pi^- e \nu$ and
$K^+ \to \pi^+ \pi^0 \pi^0$  decays, the evaluation
of isospin breaking corrections are ongoing theoretical 
efforts.
 
\section{Summary}
From highest to lowest energies, a range of phenomena induced
by the strong interaction are being explored. All are important
in order to arrive at a complete picture of QCD. 
Much headway has been made
in terms of precision measurements, while many observations
remain unexplained.

\ack
I would like to thank the conference organizers for their efforts to
make this an interesting and successful conference.
I also thank my colleagues on the various different experiments
for their input and useful discussion. 
I gratefully acknowledge support by the National Science
Foundation under contract NSF PHY-0202078.


\section*{References}
\begin{thebibliography}{30}
\bibitem{pentaquarks} 
  D. Ozerov [H1 Collaboration], 
  {\it ``Search for Baryonic Resonances at HERA''},
  this conference. 

\bibitem{pdg06} 
  W.~M.~Yao {\it et al.}  [Particle Data Group],
  J.\ Phys.\ G {\bf 33}, 1 (2006) and 2007 partial update for edition 2008.

\bibitem{e385:psi2Swidth} 
  M.~Andreotti {\it et al.}  [Fermilab E835 Collaboration],
  Phys.\ Lett.\  B {\bf 654}, 74 (2007)
  [arXiv:hep-ex/0703012].

\bibitem{cleo:chicJ_4h}
  D. Cassel [CLEO Collaboration], 
  {\it ``CLEO-c charmonium results'',}
  this conference.

\bibitem{bes:X1576} 
  M.~Ablikim {\it et al.}  [BES Collaboration],
  Phys.\ Rev.\ Lett.\  {\bf 97}, 142002 (2006)
  [arXiv:hep-ex/0606047].

\bibitem{bodwin:CcbarInBDecay}
  G.~T.~Bodwin, E.~Braaten and G.~P.~Lepage,
  Phys.\ Rev.\  D {\bf 51}, 1125 (1995)
  [Erratum-ibid.\  D {\bf 55}, 5853 (1997)]
  [arXiv:hep-ph/9407339].

\bibitem{babar:BToh_cAndEta_c} 
  B. Aubert {\it et al.}  [BABAR Collaboration],
  SLAC-PUB-12665 and arXiv:0707.2843 [hep-ex] (EPS07 conference submission).

\bibitem{bes:2-5GevScan} 
  M.~Ablikim {\it et al.}  [BES Collaboration],
  arXiv:0705.4500 [hep-ex]. 

\bibitem{belle:DD2star}
  G.~Pakhlova {\it et al.}  [Belle Collaboration],
  arXiv:0708.3313 [hep-ex] (submitted to Phys. Rev. Letters).

\bibitem{belle:ddstar_dstardstar} 
  K.~Abe {\it et al.}  [Belle Collaboration],
  Phys.\ Rev.\ Lett.\  {\bf 98}, 092001 (2007)
  [arXiv:hep-ex/0608018].

\bibitem{cleo:dstardpi} S. Stone [CLEO Collaboration], 
  {\it ``Measurements of Hadronic Cross Sections at CLEO'',}
  this conference.

\bibitem{belle:y4260} 
  C.~Z.~Yuan {\it et al.}  [Belle Collaboration],
  Phys.\ Rev.\ Lett.\  {\bf 99}, 182004 (2006)
  [arXiv:0707.2541 [hep-ex]]. 

\bibitem{belle:psi2Spipi} 
  X.~L.~Wang {\it et al.}  [Belle Collaboration],
  Phys.\ Rev.\ Lett.\ {\bf 99}, 142002 (2007) 
  [arXiv:0707.3699 [hep-ex]]. 

\bibitem{belle:y3940}
  K.~Abe {\it et al.}  [Belle Collaboration],
  Phys.\ Rev.\ Lett.\  {\bf 94}, 182002 (2005)
  [arXiv:hep-ex/0408126].

\bibitem{babar:y3940} 
  G. Cibinetto [BaBar Collaboration], 
  {\it ``Quarkonium spectroscopy and searches for new states at BaBar'',}
  this conference.

\bibitem{belle:x4160}
  K.~Abe {\it et al.}  [Belle Collaboration],
  BELLE-CONF-0705 and arXiv:0708.3812 [hep-ex] (EPS07 conference submission).

\bibitem{belle:ds1_2536} 
  V. Balagura [Belle Collaboration], 
  {\it ``Decays of charmed hadrons at Belle'',}
  this conference; see also arXiv:0709.4184 (submitted to Phys. Rev. D).

\bibitem{baryon:spectroscopy} 
  A. Savoy-Navarro, 
  {\it ``Flavour Physics at Other Facilities'',}
  plenary talk at this conference. 

\bibitem{babar:baryons} 
  T. Schroeder [BaBar Collaboration], 
  {\it ``Charm spectroscopy in BaBar'',}
  this conference.

\bibitem{kloe:pi0pi0gamma} 
  F.~Ambrosino {\it et al.}  [KLOE Collaboration],
  Eur.\ Phys.\ J.\  C {\bf 49}, 473 (2007)
  [arXiv:hep-ex/0609009].

\bibitem{kloe:eta_gluonium}
  F.~Ambrosino {\it et al.}  [KLOE Collaboration],
  Phys.\ Lett.\  B {\bf 648}, 267 (2007)
  [arXiv:hep-ex/0612029].

\bibitem{kloe:gluonium}
  F.~Ambrosino {\it et al.}  [KLOE Collaboration],
  Phys.\ Lett.\  B {\bf 648}, 267 (2007)
  [arXiv:hep-ex/0612029].

\bibitem{bes:etainvis}
  M.~Ablikim {\it et al.}  [BES Collaboration],
  Phys.\ Rev.\ Lett.\  {\bf 97}, 202002 (2006)
  [arXiv:hep-ex/0607006].

\bibitem{cleo:etabr} 
  A.~Lopez {\it et al.}  [CLEO Collaboration],
  Phys. Rev. Lett. {\bf 99}, 122001 (2007)
  [arXiv:0707.1610 [hep-ex]].

\bibitem{cleo:etamass} 
  D.~H.~Miller {\it et al.}  [CLEO Collaboration],
  Phys. Rev. Lett. {\bf 99}, 122002 (2007)
  [arXiv:0707.1810 [hep-ex]].

\bibitem{kloe:eta_3pi} 
  F.~Ambrosino {\it et al.}  [KLOE Collaboration],
  arXiv:0707.2355 [hep-ex] (LP07 conference submission).

\bibitem{na48:k_3pi_and_ke4} 
  G. Lamanna [NA48 Collaboration], 
  {\it ``$\pi \pi$ scattering from $Ke4$ decays'',}
  this conference.
\end{thebibliography}
\end{document}